\documentclass[letterpaper,english,aps,preprintnumbers,nofootinbib,showpacs,prd,notitlepage]{revtex4-1}
\usepackage{amsmath}
\usepackage{makeidx}
\usepackage{float}
\usepackage{graphicx}
\usepackage[subrefformat=parens,labelformat=parens,caption=false]{subfig}
\usepackage{hyperref}
\usepackage{cancel}
\usepackage{amsfonts}
\usepackage{diagbox}

\begin{document}
\def\({\left(}
\def\){\right)}
\def\[{\left[}
\def\]{\right]}

\global\long\def\M{\mathcal{M}}
\global\long\def\O{\mathcal{O}}
\global\long\def\GF{\text{G}_\text{F}}
\global\long\def\d{\text{d}}
\global\long\def\gnn{\text{Ps}\to \gamma \nu_\ell \bar{\nu}_\ell}
\global\long\def\pggg{\text{p-Ps}\to 3\gamma}
\global\long\def\piggg{\pi^{0}\to3\gamma}
\global\long\def\ognn{\text{o-Ps}\to\gamma\nu_{\ell}\bar{\nu}_{\ell}}
\global\long\def\oggg{\text{o-Ps}\to3\gamma}
\global\long\def\mw{m_\text{W}}
\global\long\def\mz{m_\text{Z}}
\global\long\def\me{m_\text{e}}
\global\long\def\tr{\text{Tr}}
\global\long\def\gw{g_\text{w}}
\global\long\def\nn{\nonumber\\}
\global\long\def\={&=}

\raggedbottom

\preprint{Alberta Thy 19-17}

\title{Parapositronium can decay into three photons}

\author{Andrzej Pokraka}
\email{pokraka@ualberta.ca}

\author{Andrzej Czarnecki}
\email{andrzejc@ualberta.ca}

\affiliation{Department of Physics, University of Alberta, Edmonton, Alberta,
Canada T6G 2E1}

\begin{abstract}
We provide the first explicit calculation showing that parapositronium can decay 
into three photons. While this decay is forbidden within quantum 
electrodynamics because it violates charge conjugation symmetry, it 
proceeds through the weak interaction. We compute the charged weak 
boson contribution to the photon energy spectrum and the rate of this decay.
\end{abstract}

\maketitle

\section{Introduction}
Positronium (Ps) is an atom composed of an electron and a positron. 
The ground state is the spin singlet parapositronium (p-Ps) while 
the lowest excited state is the spin triplet orthopositronium (o-Ps). 

Within pure quantum electrodynamics (QED), p-Ps can decay only 
into an even number of photons and o-Ps only into an odd number 
of photons. This is because QED is invariant under charge conjugation 
symmetry ($C$-symmetry).

With the inclusion of $C$-violating weak interactions, Ps gains access 
to new photonic decay modes. The simplest one being $\pggg$ (the 
$\text{o-Ps}\to2\gamma$ is forbidden by the Landau-Yang theorem 
\cite{Landau:1948kw,Yang:1950rg}). While Ref.~\cite{Bernreuther:1981ah} 
argues that this decay is possible and estimates its branching ratio,
their statement has not yet been supported by an explicit calculation. 

We compute the $W$-boson contribution to the $\pggg$ decay and thus 
demonstrate that it does indeed occur.
In Sec.~\ref{sec:TotalAmp} we calculate the amplitude generated by $W$-boson 
loops to order $\O\(\me^6/\mw^6\)$ where $\me$ is the mass of the electron and 
$\mw$ is the mass of the $W$-boson. In Sec.~\ref{sec:decayrate} we use the result 
of Sec.~\ref{sec:TotalAmp} to compute the decay rate, branching ratio and 
photon spectrum of $\pggg$ decays. We conclude in Sec.~\ref{sec:conclusions}.

\section{Decay amplitude \label{sec:TotalAmp}}
We start by noting that since the $\pggg$ decay violates $C$-symmetry, 
it must also violate parity in order to conserve $CP$. Therefore, the final 
state must be composed of three spatially symmetric 
photons with vanishing total angular momentum -- identical to 
that of the neutral pion decay into three photons ($\pi^0\to3\gamma$) 
\cite{Dicus:1975cz}. 

The $\pggg$ decay amplitude can be written
as a sum of tensors made up of the external momentum multiplied by
scalar functions, $F_i$, called form factors. The form factors are functions
of the only available scalars: $x, y,z$ where $x,y,z = k^0_{1,2,3}/\me$ 
and $k_i$ is the 4-momentum of the $i^\text{th}$ 
photon in the final state. 

Following Ref.~\cite{Dicus:1975cz}, we the use gauge invariance of the photon field 
to construct the general form of the amplitude (detailed in Appendix \ref{sec:amp})
\begin{equation}
	\label{eq:amp with ep}
	\mathcal{M}  = \epsilon^*_{\mu_1} \epsilon^*_{\mu_2} \epsilon^*_{\mu_3} 
				\mathcal{M}^{\mu_1\mu_2\mu_3} (k_1,k_2,k_3)
\end{equation} 
where 
\begin{align}
	\mathcal{M}^{\mu_{1}\mu_{2}\mu_{3}}(k_{1},k_{2},k_{3}) 
	& = \left(
			k_{1}^{\mu_{3}}
			- k_{2}^{\mu_{3}}\frac{k_{1}\cdot k_{3}}{k_{2}\cdot k_{3}}
		\right)
		\left(k_{1}^{\mu_{2}}k_{2}^{\mu_{1}}-k_{1}\cdot k_{2}g^{\mu_{2}\mu_{1}}\right) 
		F_{1} (x,y,z)
	\nonumber \\
 	& +\left(
			k_{1}^{\mu_{2}}
			-k_{3}^{\mu_{2}} \frac{k_{1}\cdot k_{2}}{k_{2}\cdot k_{3}}
		\right)
		\left(k_{3}^{\mu_{1}}k_{1}^{\mu_{3}}-k_{3}\cdot k_{1}g^{\mu_{1}\mu_{3}}\right)
 		F_{2} (x,y,z)
	\nonumber \\
 	& +\left(
			k_{2}^{\mu_{1}}
			-k_{3}^{\mu_{1}}
			\frac{k_{1}\cdot k_{2}}{k_{1}\cdot k_{3}}
		\right) 
		\left(k_{3}^{\mu_{2}}k_{2}^{\mu_{3}}-k_{2}\cdot k_{3}g^{\mu_{2}\mu_{3}}\right)
		F_{3} (x,y,z)
	\nonumber \\
 	& +\big[
			k_{2}^{\mu_{3}}
			\left(k_{1}^{\mu_{2}}k_{3}^{\mu_{1}}-k_{1}\cdot k_{3}g^{\mu_{2}\mu_{1}}\right)
			-k_{1}^{\mu_{3}}
			\left(k_{3}^{\mu_{2}}k_{2}^{\mu_{1}}-k_{3}\cdot k_{2}g^{\mu_{2}\mu_{1}}\right)
	\nonumber \\
 	& \quad
			+ k_{2}^{\mu_{1}}k_{1}\cdot k_{3}g^{\mu_{3}\mu_{2}}
			+ k_{3}^{\mu_{2}}k_{1}\cdot k_{2}g^{\mu_{3}\mu_{1}}
			- k_{1}^{\mu_{2}}k_{2}\cdot k_{3}g^{\mu_{3}\mu_{1}}
			- k_{3}^{\mu_{1}}k_{1}\cdot k_{2}g^{\mu_{3}\mu_{2}}
		\big] F_{4}(x,y,z).
\label{eq:amp}
\end{align}
and the $\epsilon_i$ are the photon polarizations.
Furthermore, Bose symmetry places restrictions on the form factors such
that only $F_1$ and $F_4$ need to be calculated,
\begin{align}
F_{2}(x,y,z) & =F_{1}(x,z,y),
\label{eq:F2F1}
\\
F_{3}(x,y,z) & =F_{1}(y,z,x).
\label{eq:F3F1}
\end{align}
The form factors $F_1$ and $F_4$ can be projected out from the total amplitude  
by contracting $\mathcal{M}^{\mu_{1}\mu_{2}\mu_{3}}(k_{1},k_{2},k_{3}) $ 
with suitable tensors (Appendix \ref{sec:amp}). 

To lowest order in perturbation theory, $\pggg$ proceeds through
one-loop processes in electroweak theory. It is especially convenient
to employ the  the non-linear  
renormalizable gauge ($R_\xi$) \cite{Fujikawa:1973qs,Gavela:1981ri},
which eliminates  the three point Goldstone-$W$-photon 
vertex, reducing the number of  
diagrams that need to be computed. In this formalism, 
\begin{align}
	\mathcal{L}_\text{Gauge-fix} 
	& = 	-\frac{1}{2\alpha}(\partial^\mu A_\mu)^2
		-\frac{1}{2\eta} \(\partial^\mu Z_\mu - \eta m_\text{Z} \chi\)^2
	\nn
		& \quad - \frac{1}{\xi} \(\partial^\mu W^{+}_\mu - i \xi \mw \phi^+ - ig A^3_\mu {W^+}^\mu \)
	                              \(\partial^\mu W^{-}_\mu + i \xi \mw \phi^- + ig A^3_\mu {W^-}^\mu \),
\end{align}
where $\alpha,\eta$ and $\xi$ are gauge parameters. The Feynman rules 
in this gauge are listed in Appendix \ref{sec:Rules}. 

To calculate the amplitude, the electron and positron are approximated to be at 
rest with four-momentum $p=(m,\mathbf{0})$ and the p-Ps projection operator 
$\Psi_\text{p-Ps} = (1+\gamma^0)\gamma^5/2\sqrt2$ \cite{Czarnecki:1999mw} is 
used to project out the correct spin configuration of the electron and positron.  

Similar to the $\piggg$ and $\ognn$ decays, the diagrams that contribute to
$\pggg$ must contain an axial-vector interaction in the trace along the fermion 
line in order to break $C$-symmetry \cite{Dicus:1975cz,Pokraka:2016jgy}. 
At the one-loop level, this restricts the relevant diagrams to those that do not 
contain virtual photons.

\begin{figure}[t] 
	\centering
	\subfloat[]{\label{fig:W1}
		\includegraphics[width=.23\textwidth]{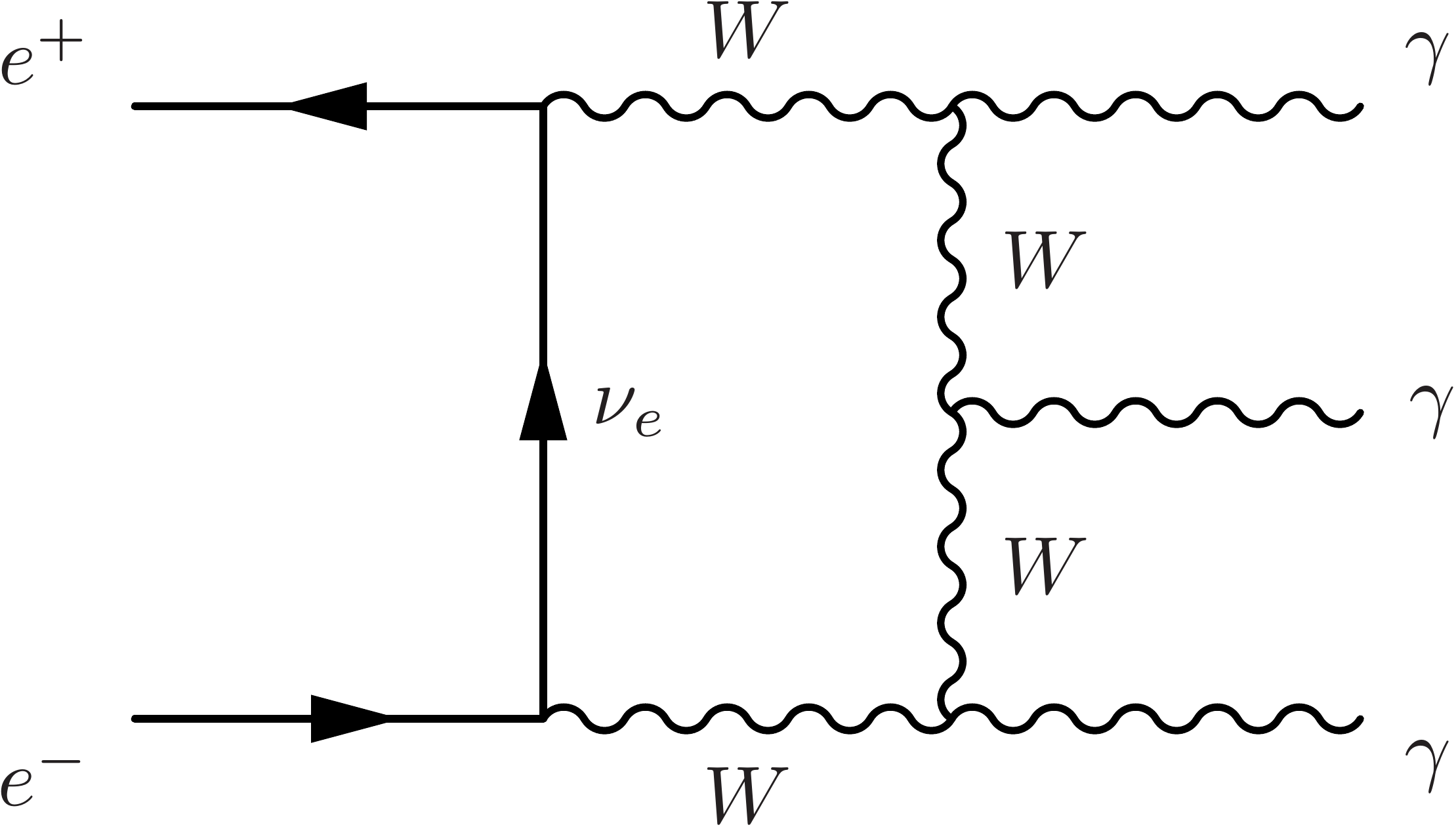}
	}
	\subfloat[]{\label{fig:W2}
		\includegraphics[width=.23\textwidth]{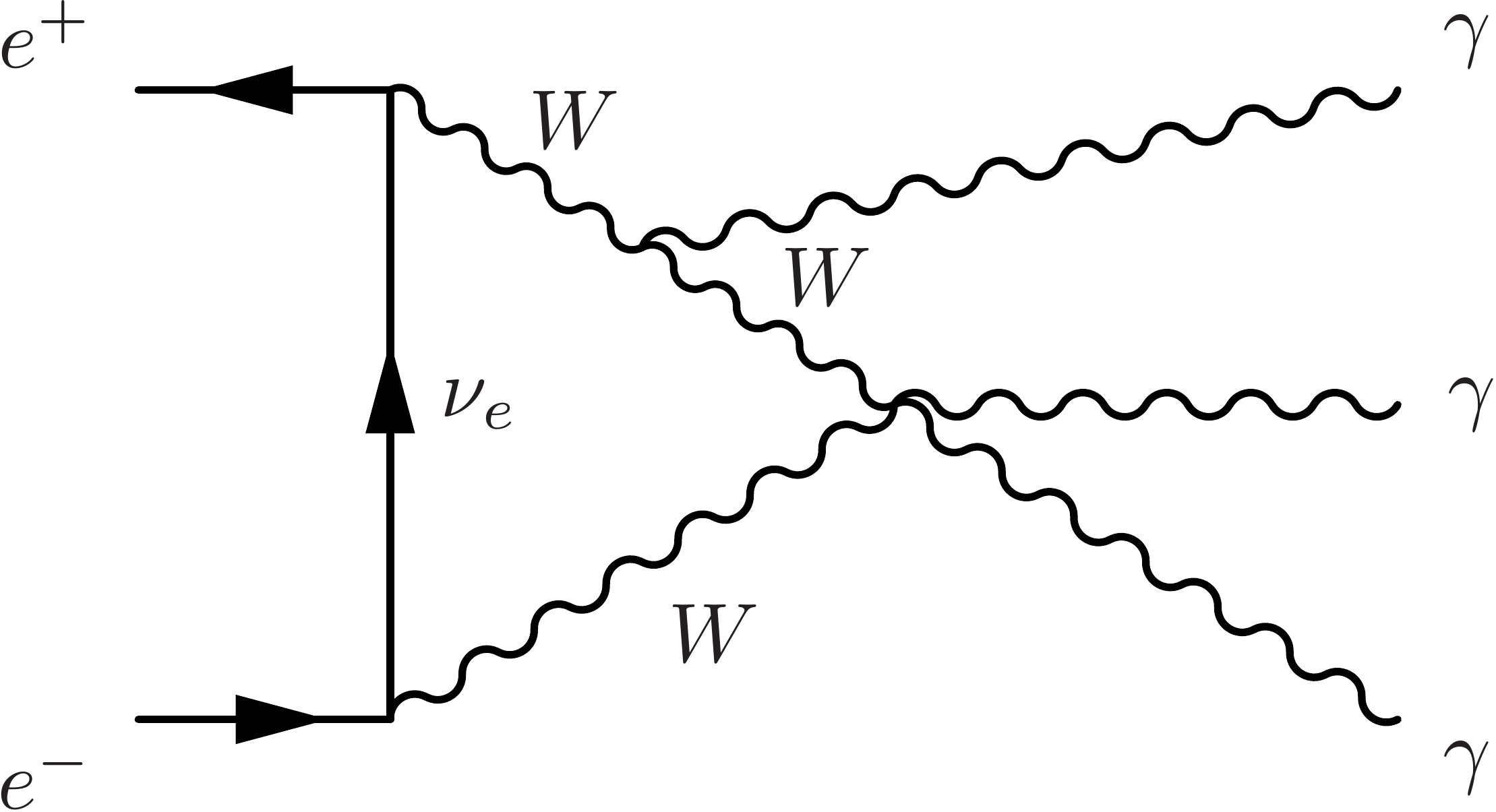}
	}
	\subfloat[]{\label{fig:W3}
		\includegraphics[width=.23\textwidth]{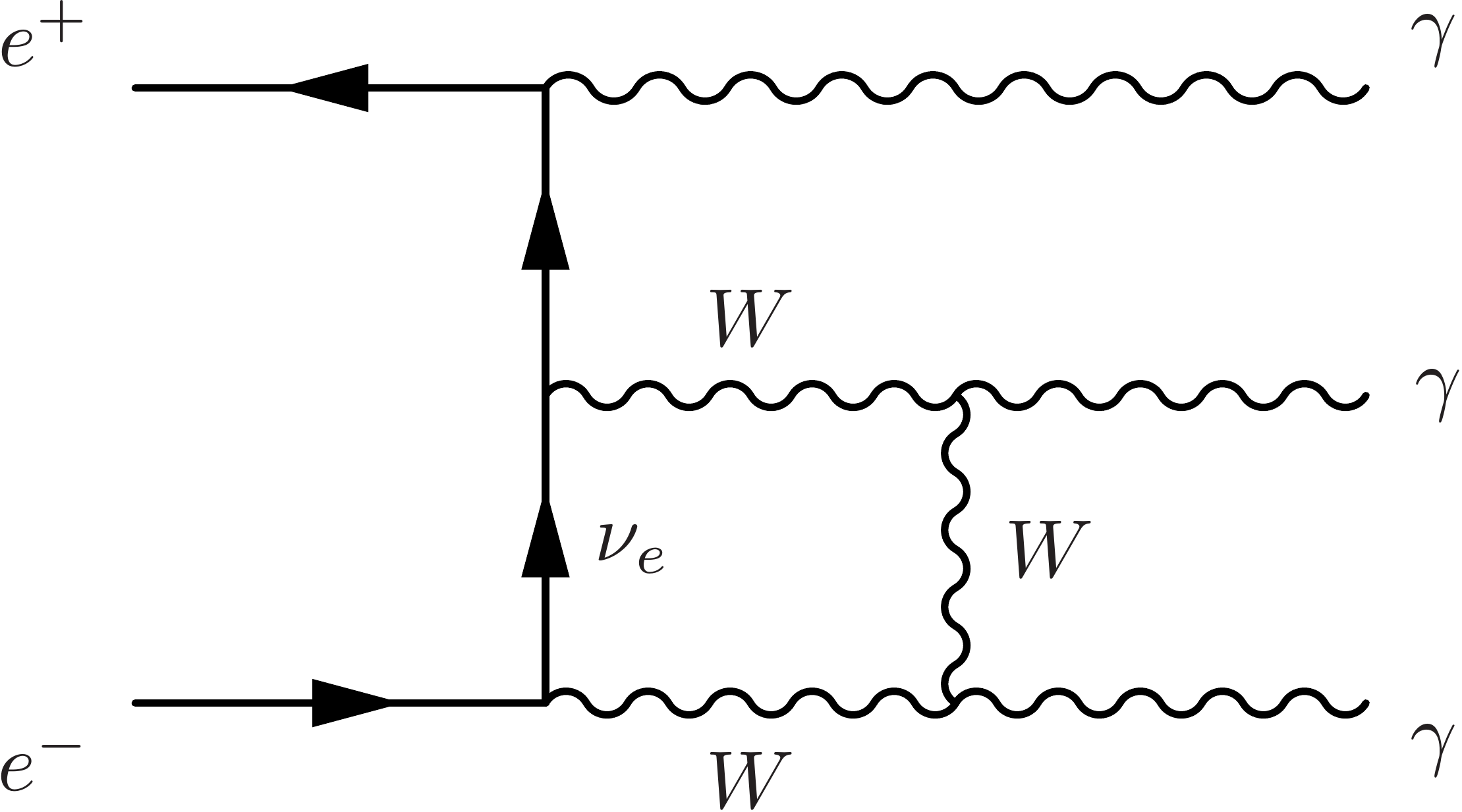}
	}
	\subfloat[]{\label{fig:W4}
		\includegraphics[width=.23\textwidth]{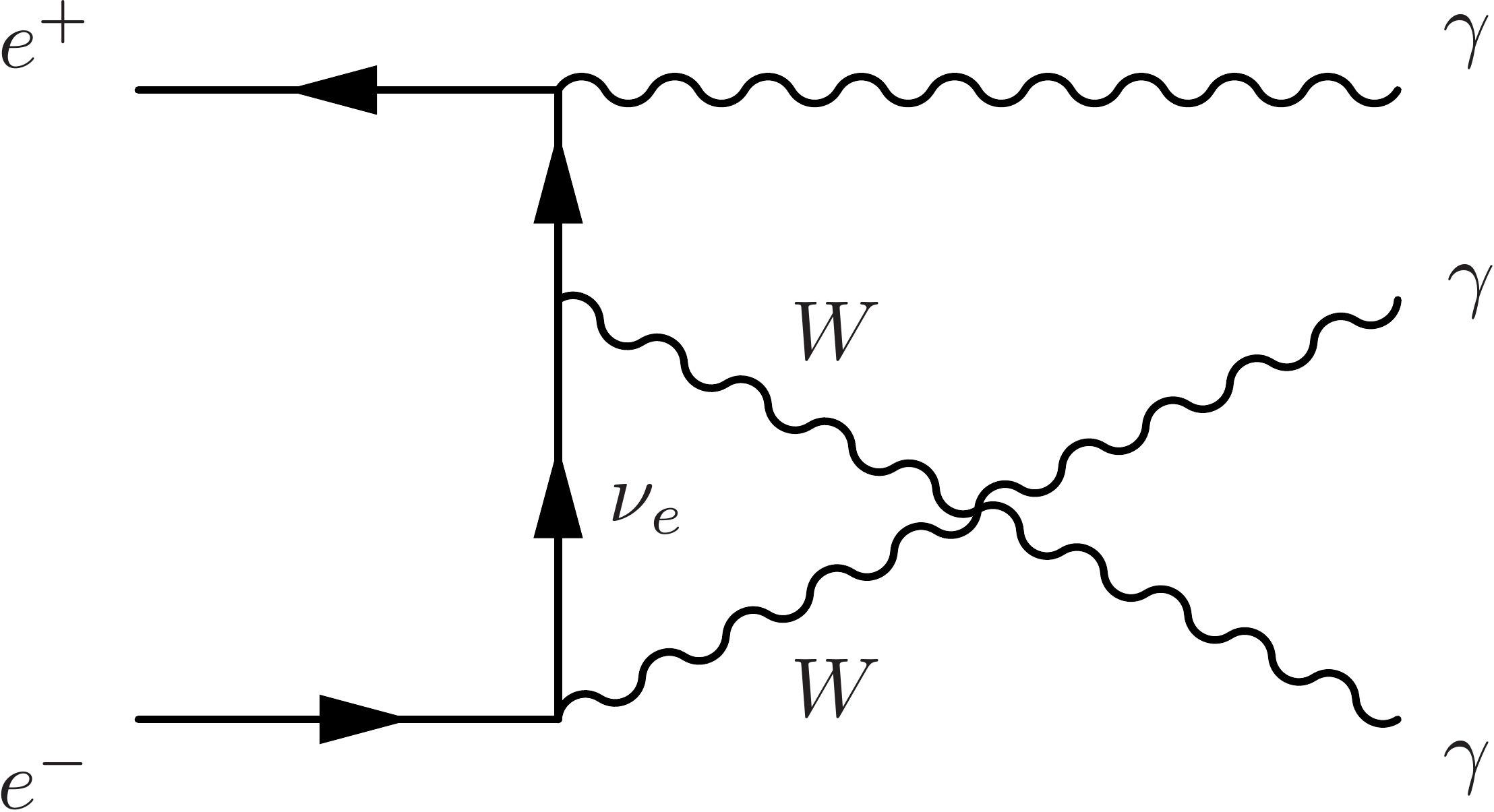}
	}
	
	\subfloat[]{\label{fig:W5}
		\includegraphics[width=.23\textwidth]{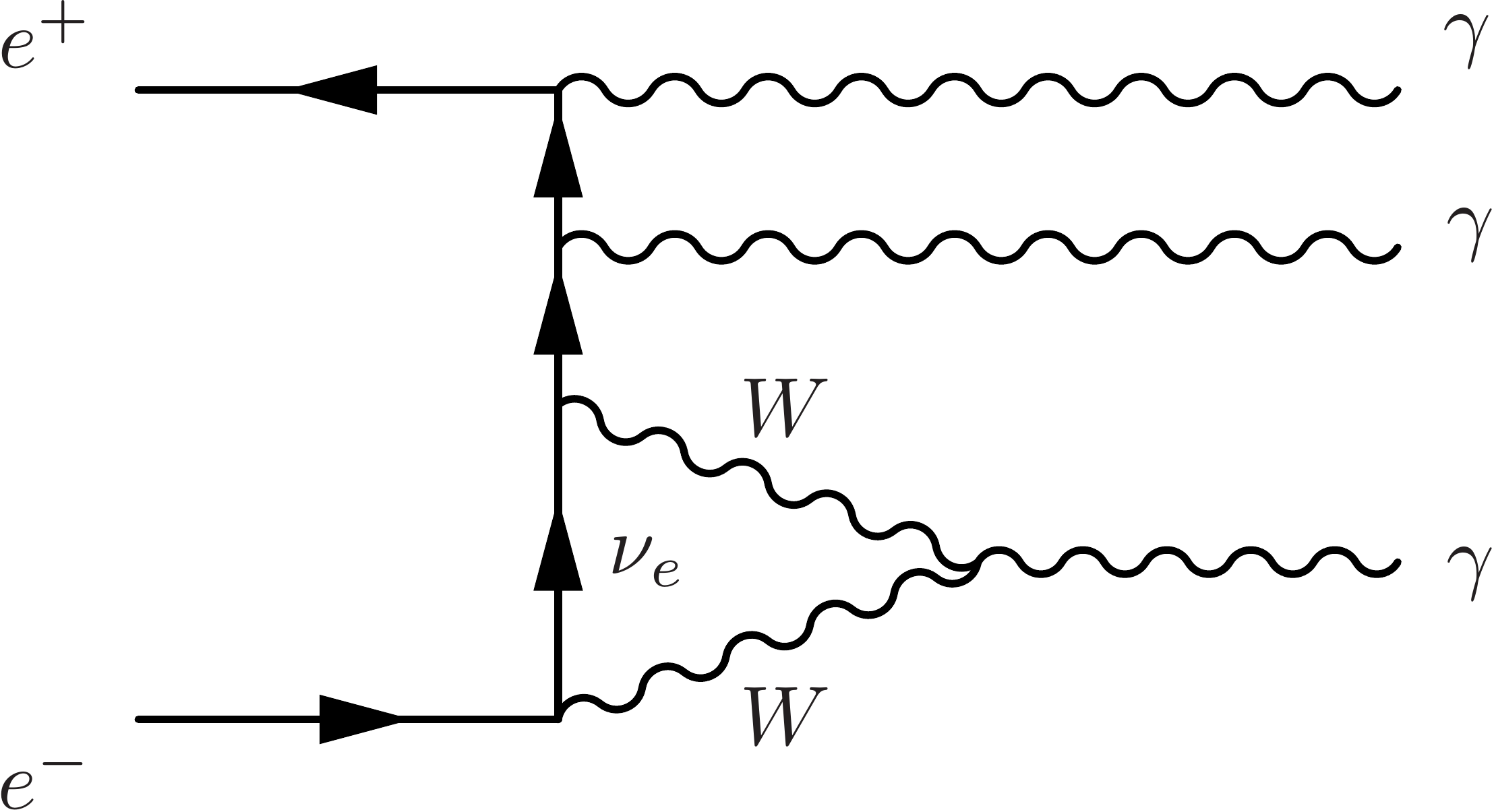}
	}
	\subfloat[]{\label{fig:W6}
		\includegraphics[width=.23\textwidth]{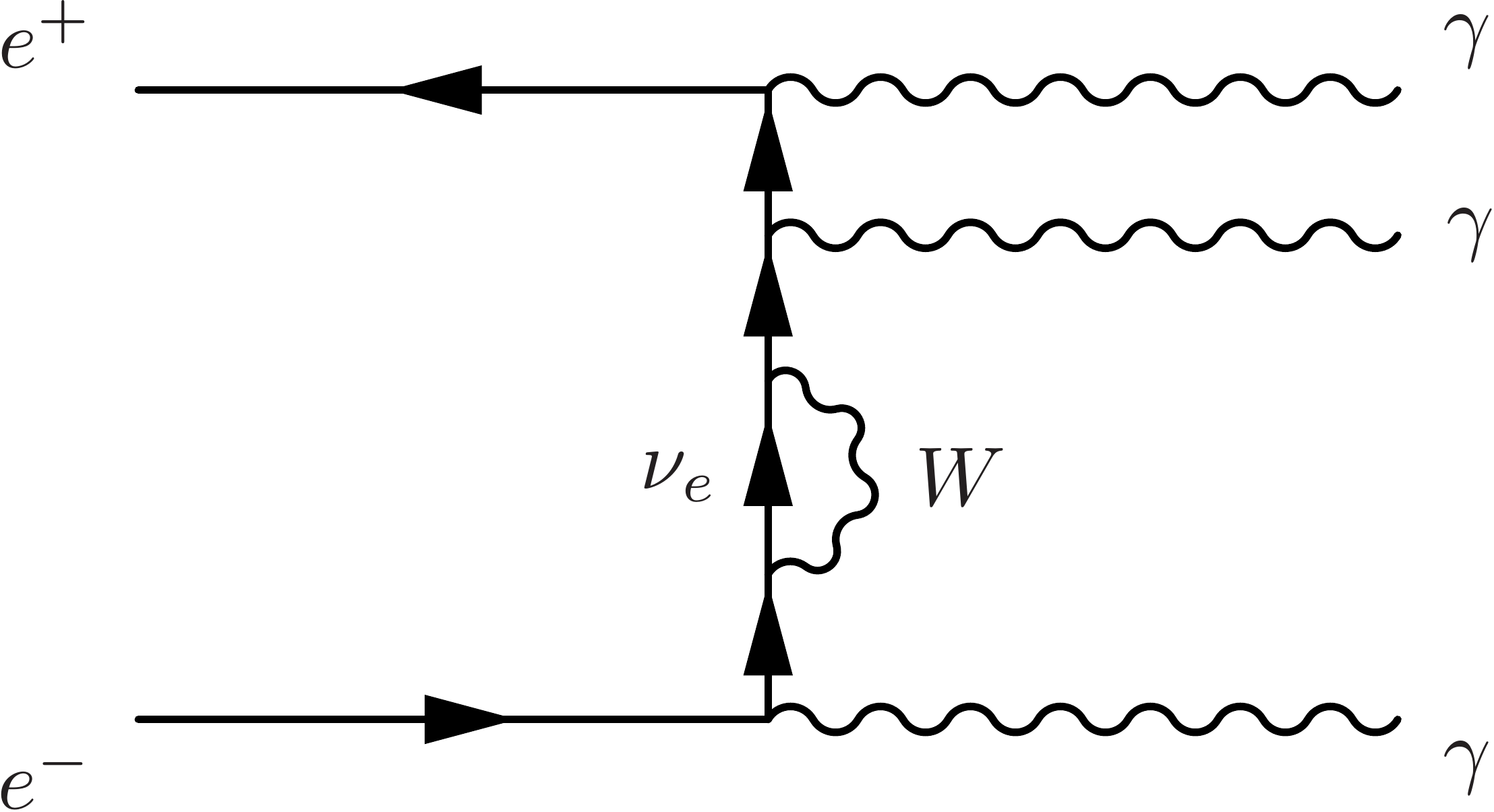}
	}
	\subfloat[]{\label{fig:W7}
		\includegraphics[width=.23\textwidth]{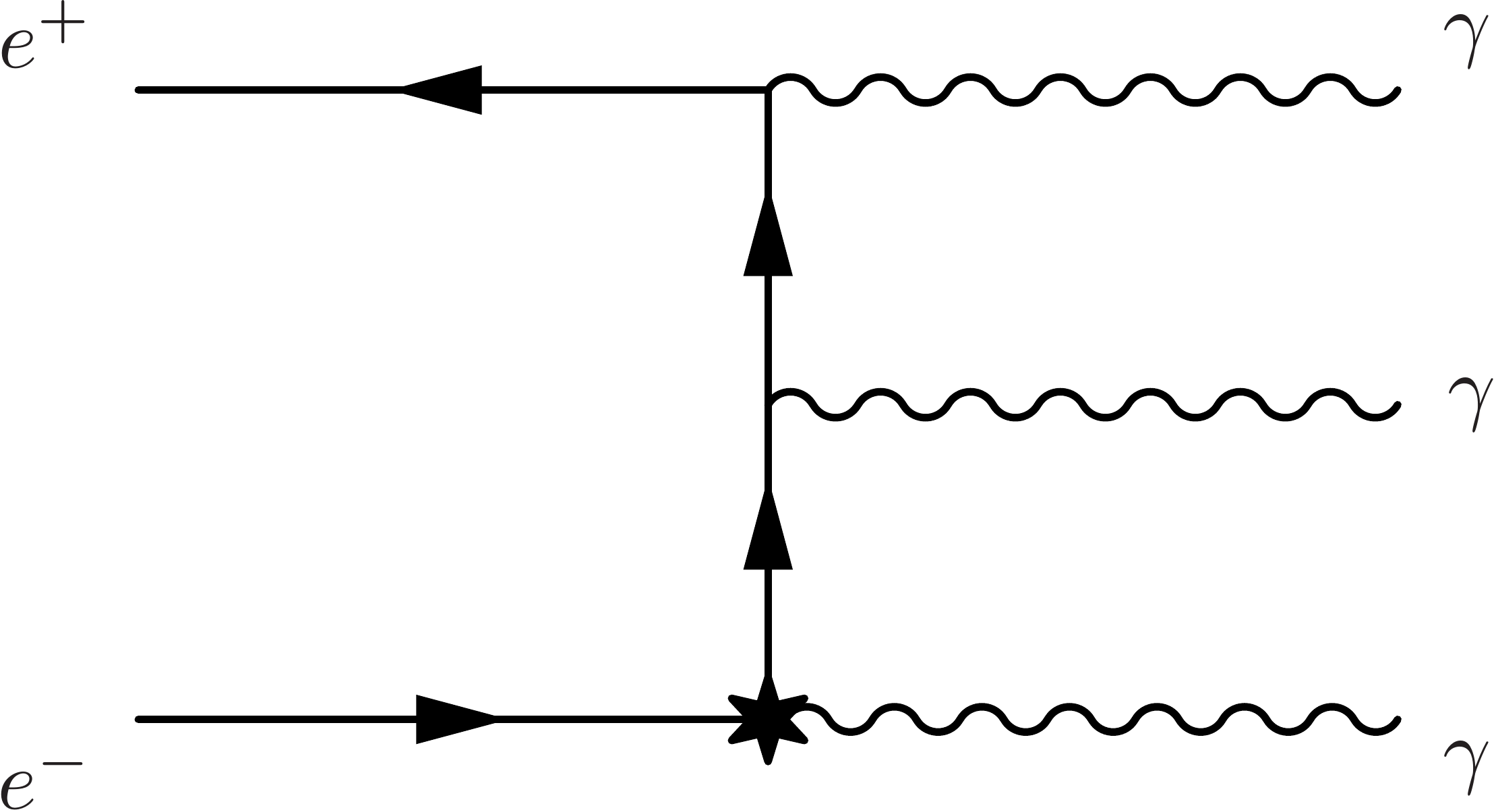}
	}
	\subfloat[]{\label{fig:W8}
		\includegraphics[width=.23\textwidth]{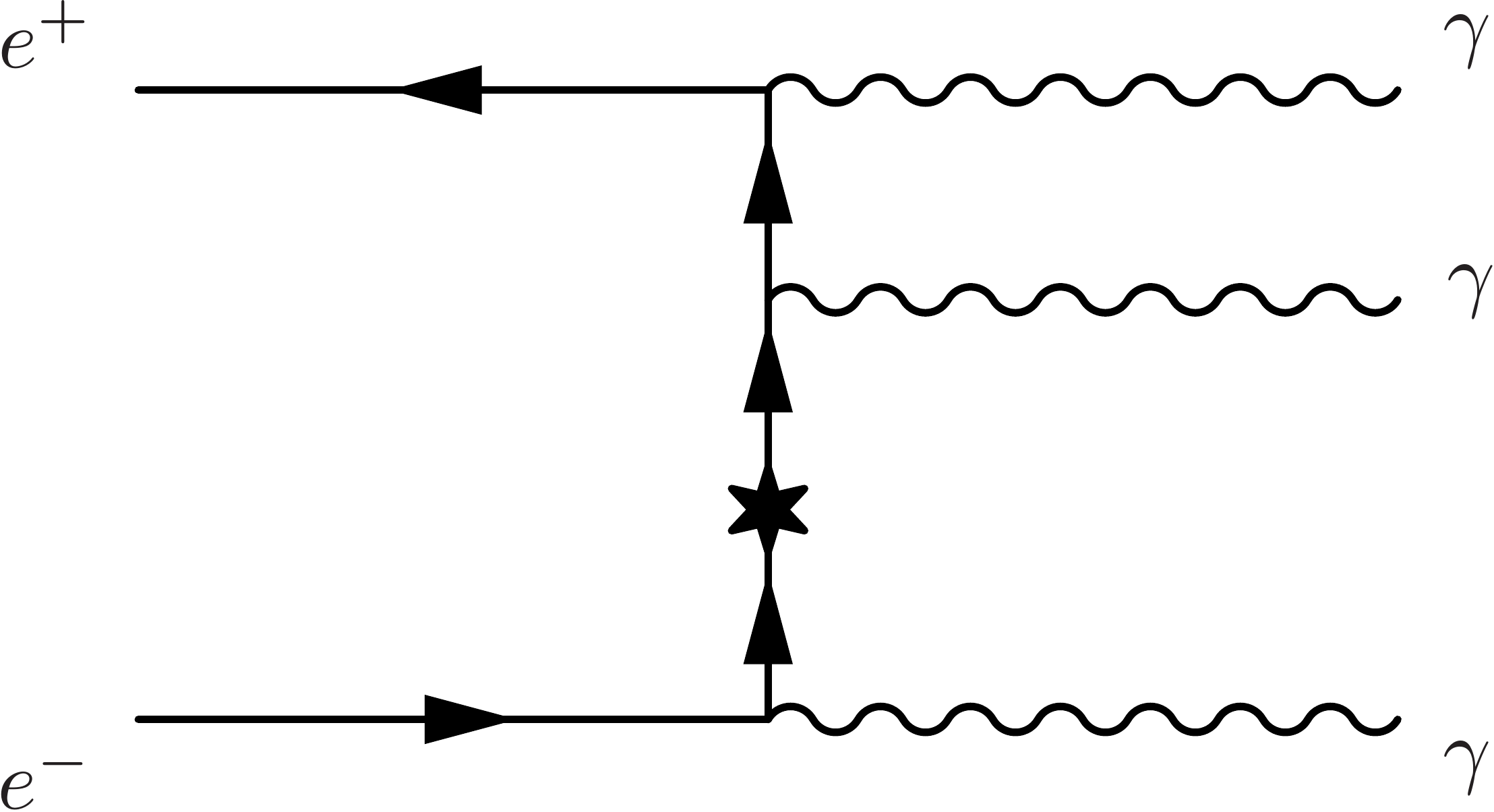}
	}
		
	\caption{\label{fig:diagrams}
		The $W$-boson contributions to the $C$-violating amplitude of $\pggg$.
		In addition to Figs.~\protect\subref{fig:W1}-\protect\subref{fig:W6}
		there are the analogous diagrams where the $W$-bosons are 
		replaced by charged Goldstone bosons $\phi^\pm$.
		Figs.~\protect\subref{fig:W7} and \protect\subref{fig:W8}, 
		are counterterm diagrams that remove the divergences of 
		Figs. \protect\subref{fig:W5} and \protect\subref{fig:W6}.
		}
\end{figure}

Since our goal is to ascertain whether the $\pggg$ decay is possible, we can 
examine either the $W$- or $Z$-boson contributions to this decay. 
For simplicity, we focus our attention, in this paper, on diagrams containing $W$-bosons; 
these diagrams are given in Fig.~\ref{fig:diagrams}. 
Counterterm diagrams \ref{fig:diagrams}\protect\subref{fig:W7}  and 
\ref{fig:diagrams}\protect\subref{fig:W8} (Appendix \ref{sec:renormalization})
cancel the divergences of Figs.~\ref{fig:diagrams}\protect\subref{fig:W5} and 
\ref{fig:diagrams}\protect\subref{fig:W6}.

As a consistency check, we keep the gauge parameter of the $W$-bosons 
and show that the form factors in the total amplitude are gauge 
independent. Namely, 
\begin{equation} \label{eq:FF}
	F_{i} (x,y,z=2-x-y)
		=\frac{7\GF e^{3}m_{e}}{360 \pi^2 \mw^{4}} (1-x) f_{i}(x,y)
\end{equation}
where
\begin{align}
	f_{1} & =\frac{1-y}{x}-\frac{1-x}{y},
	\label{eq:f1}
	\\
	f_{2} & = (1-y) \left(\frac{1}{2-x-y}-\frac{1}{x}\right),
	\\
	f_{3} & = (1-y) \left(\frac{1}{2-x-y}-\frac{1}{y}\right),
	\\
	f_{4} & = 0.
	\label{eq:f4}
\end{align}
Here, $\GF\simeq 1.166 \cdot 10^{-5}/\text{GeV}^2$ is the Fermi coupling constant 
\cite{Marciano:1999ih} and $e>0$ is the electric charge of a proton. Substituting 
these form factors into \eqref{eq:amp} yields the $W$-boson contribution to 
total $\pggg$ decay amplitude. The form factors
\eqref{eq:f1}-\eqref{eq:f4} have corrections of order $\mathcal{O}(\me^2/\mw^2)$ 
that we ignore. Their leading order terms are remarkably simple compared to the 
one-loop form factors for the analogous decay of o-Ps \cite{adkins:2010}. This is because the 
$W$-loop is essentially point-like relative to the other distance
scales in Ps. 

As mentioned above, both $W$- and $Z$-bosons can contribute to this decay. 
Diagrams containing virtual $Z$-bosons are proportional to the factor
$1-4\sin^2\theta_\text{W}$ from the vector coupling in the $Ze^+e^-$ 
vertex (where $\theta_{\text W}$ is the weak mixing angle and 
$\sin^2\theta_{\text W} \approx 0.238$ is numerically close to $1/4$ 
\cite{Czarnecki:2005pe}). Despite this suppression, the $Z$-boson
diagrams likely dominate because the $Z$-boson loops depend on two 
mass scales: $\me$ and $\mz$. The large distance scale related to $\me$ 
can enhance the $Z$-boson mediated decay relative to the $W$-boson 
mediated decay. 


Explicit calculation of the $Z$-boson contribution is outside the scope 
of this paper. Since the $Z$-boson loops depend on two mass scales, 
evaluation of $Z$-loops is significantly more difficult than the W-loops.


\section{Decay Rate and Photon Spectrum  \label{sec:decayrate}}
With the results of the previous section, it is simple to obtain the decay rate and 
photon spectrum. Using \eqref{eq:amp} with \eqref{eq:FF}, we obtain
\begin{align} \label{eq:rate}
	\Gamma(\pggg;W\text{-loops}) & = \frac{(14700\pi^2-145081) \GF^2 \alpha^6 \me^{13}}{58320000\pi^5 \mw^8}
		\approx 3.5 \cdot 10^{-67} \text{s}^{-1}. 
\end{align}
The corresponding branching ratio is
\begin{align} \label{eq:BR}
	\text{Br}(\pggg;W\text{-loops}) 
	= \frac{\Gamma(\text{p-Ps}\to3\gamma;W\text{-loops only})}{\Gamma(\text{p-Ps}\to2\gamma)}
	\approx 4.4 \cdot 10^{-77}.
\end{align} 
Comparing with the estimated branching ratio of \cite{Bernreuther:1981ah},
\begin{equation}
	\alpha \left[ \GF \me^2 (1-4\sin^2\theta_W) \right]^2 \approx 10^{-27},
\end{equation}
we see that the $W$-loop contribution is tiny.  Our qualitative
understanding of this suppression
is as follows. Consider the annihilation of $e^-e^+$ at high energies (centre of mass
energies greater than about 1 GeV). In this energy regime, the mass of the electron/positron 
may be ignored and mass of the $W$-boson becomes the only relevant mass scale. 
Photon gauge invariance requires three powers of $k_i$ in the amplitude, each of which 
must be divided by the only available mass: $\mw$. When transitioning from the 
high energy to the low energy regime, the $W$-boson mass dividing the
$k_i$'s in the amplitude cannot be replaced by the electron mass. Of course, 
in a low energy process, the electron mass becomes important and may give rise to 
additional contributions, but they will be no less suppressed by the $W$-boson mass. 

\begin{figure}[t] 
	\centering
	\includegraphics[width=.45\textwidth]{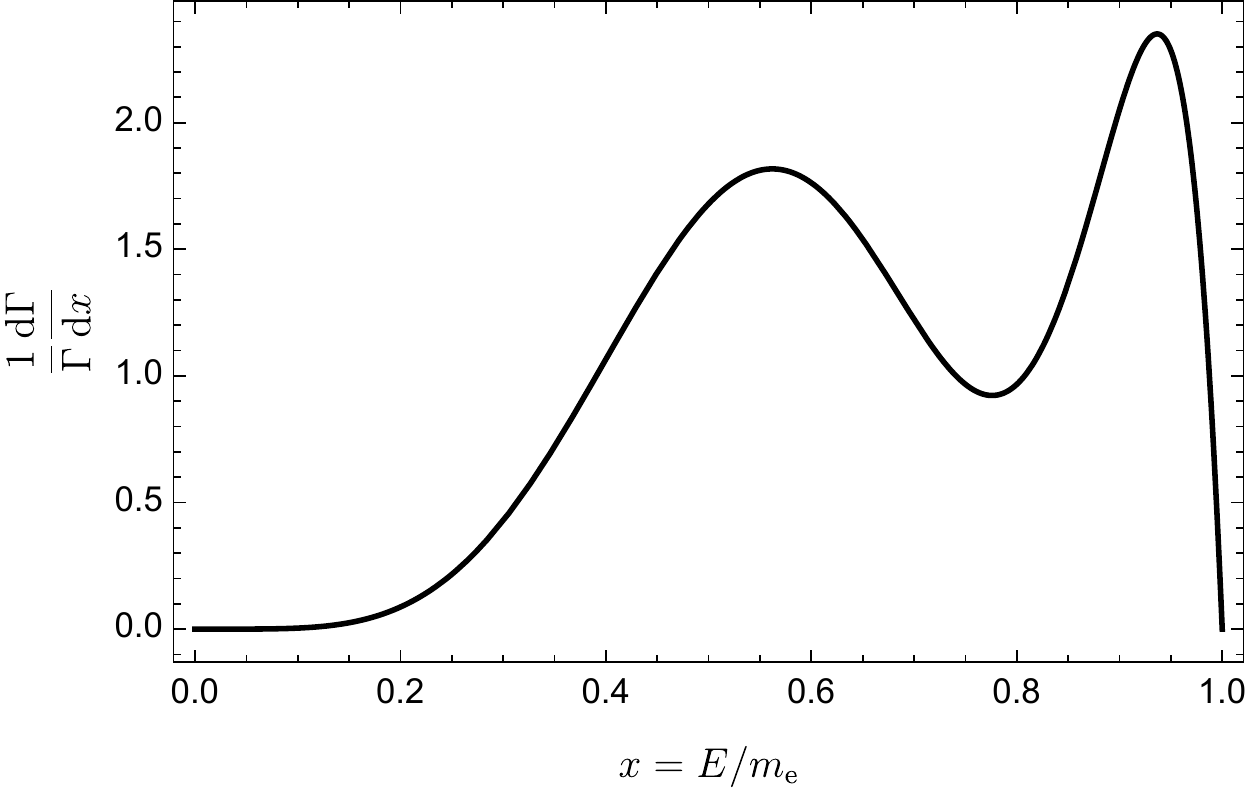}
	\caption{\label{fig:spectrum}
		Plot of the $W$-boson contribution to the 
		$\pggg$ photon spectrum \eqref{eq:spectrum}.  
		Here, $x=E/\me$ is the energy of a photon, $E$, divided by the 
		electron mass.
		}
\end{figure}

The corresponding photon spectrum is calculated to be
\begin{align} \label{eq:spectrum}
	\frac{1}{\Gamma}\frac{\d \Gamma}{\d x} = &
	\frac{70 (1-x)} {14700 \pi^2 - 145081}
	\bigg[
		-840 + 7700x^2 - 14560x^3 + 10192x^4
	\nonumber \\ \qquad & 
		-2489x^5
		-\frac{840 (2+6x-6x^2+x^3) (1-x)^4 \ln (1-x)}{(2-x) x}
	\bigg]
\end{align}
where $x$ is the energy of one of the photons divided by the electron mass
and $\Gamma=\Gamma(\pggg;W\text{-loops})$ is given in \eqref{eq:rate}. Equation
\eqref{eq:spectrum} is plotted in Fig.~\ref{fig:spectrum}. 

When one photon has maximal energy ($x=1$), the other two photons must move 
collinearly and in the opposite direction. However, this configuration does not conserve 
angular momentum and thus the photon spectrum \eqref{eq:spectrum} vanishes at $x=1$. 
This is in contrast with the spectrum of the orthopositronium decay into $3\gamma$, which 
reaches its maximum at $x =1$.
Fig.~\ref{fig:spectrum} also shows the suppression of low-energy photon emission by a neutral 
system. In the low energy limit the photon spectrum is of $\mathcal{O}(x^5)$ in agreement
with Low's theorem \cite{Low:1958sn}.

\section{Conclusions \label{sec:conclusions}}
We have presented the first calculation that explicitly demonstrates
that  parapositronium decays into an odd number of photons. 
Positronium has been used to test $C$-symmetry by looking for 
forbidden QED decay modes such as $\pggg$ and 
$\text{o-Ps}\to4\gamma$ \cite{Moskal:2016moj,Gninenko:2006sz,Mills:1967yok}. 
Together with the previously published estimate of the $Z$-loops
\cite{Bernreuther:1981ah}, our calculation demonstrates that standard model 
contributions to $\pggg$ decays are far smaller than conceivable experimental 
sensitivities. Any detection of $\pggg$, in the near future would be a signal of new 
physics  \cite{Caravati:2002ax,Grosse:2001ei}.

\section*{Aknowledgments}
This research was supported by the Natural Sciences and Engineering Research 
Council of Canada (NSERC).


\appendix
\section{Amplitude structure \label{sec:amp}}
This appendix summarizes the construction of the amplitude presented in \cite{Dicus:1975cz}.
We start by defining the matrix element for $\pggg$ by
\begin{equation}
	\M (k_{1},k_{2},k_{3})=
		\epsilon_{\mu_{1}}^{*} \epsilon_{\mu_{2}}^{*} \epsilon_{\mu_{3}}^{*} 
			\mathcal{M}^{\mu_{1}\mu_{2}\mu_{3}}(k_{1},k_{2},k_{3})
\end{equation}
where $\epsilon_{i}$ are the polarizations of the external photons. 
The most general third rank tensor constructed out of the external
photon momentum contains $27$ terms of the form 
$k_i^{\mu_1} k_j^{\mu_2} k_k^{\mu_3}$ and
9 terms of the form $k_l^{\mu_i} g^{\mu_j \mu_k}$.
Since the amplitude will be contracted with 
$\epsilon_{\mu_3}^* \epsilon_{\mu_2}^* \epsilon_{\mu_1}^*$
we can identify $k_{i}^{\mu_{i}}=0$. This reduces the number of terms
to 14 possible terms: 8 of the type $k_{i}^{\mu_{1}}k_{j}^{\mu_{2}}k_{k}^{\mu_{3}}$ 
and 6 of the type $k_{l}^{\mu_{i}}g^{\mu_{j}\mu_{k}}$. Gauge invariance, 
\begin{equation}
k_{3\mu_{3}}\M^{\mu_{3}\mu_{2}\mu_{1}}=k_{2\mu_{2}}\M^{\mu_{3}\mu_{2}\mu_{1}}=k_{1\mu_{1}}\M^{\mu_{3}\mu_{2}\mu_{1}}=0,
\end{equation}
then restricts the structure of the amplitude to that of equation \eqref{eq:amp}.

To project out the form factors we need the projection tensors. The $F_1$ projector is
obtained by setting 
\begin{align}
	F_1 & = - \frac{k_2 \cdot k_3}{2 (k_1 \cdot k_3) (k_1 \cdot k_2)^3}
	\\
	F_2 & = F_3 = 0
	\\ 
	F_4 & = - \frac{1}{4 (k_1 \cdot k_3) (k_1 \cdot k_2)^2}
\end{align}
in equation \eqref{eq:amp}. Similarly, setting 
\begin{align}
	F_1 &= - \frac{1}{4(k_1 \cdot k_3) (k_1 \cdot k_2)^2}
	\\
	F_2 &= \frac{1}{4 (k_1 \cdot k_3)^2 (k_1 \cdot k_2)}
	\\
	F_3 &= - \frac{1}{4 (k_2 \cdot k_3)^2 (k_1 \cdot k_2)}
	\\
	F_4 &= - \frac{1}{2 (k_2 \cdot k_3) (k_1 \cdot k_3) (k_1 \cdot k_2)}
\end{align}
in equation \eqref{eq:amp} yields the $F_4$ projector.

\section{Feynman rules in the non-linear $R_\xi$ gauge \label{sec:Rules}}
In this appendix, we list the Feynman rules used in our calculation.
The  propagator for the $W$-boson and Goldstone boson propagators are unchanged from 
the linear $R_\xi$ gauge and given by 
\begin{align}
	\begin{gathered}
		\includegraphics[scale=.25]{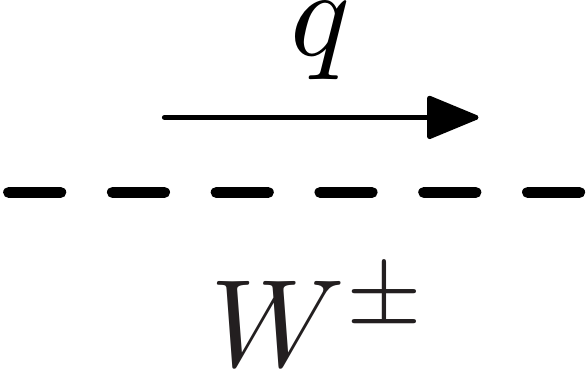}
	\end{gathered}
	\quad &= \quad \frac{-i}{q^2-\mw^2}\(g_{\mu \nu} -(1-\xi) 
		\frac{q_\mu q_\nu}{q^2-\xi m_\text{W}^2}\)
	\\
	\begin{gathered}
		\includegraphics[scale=.25]{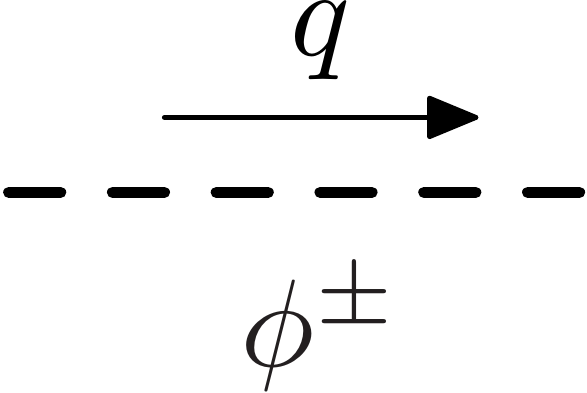}
	\end{gathered}
	\quad &= \quad \frac{i}{q^2-\xi\mw^2}
\end{align}
where $\xi$ is the gauge parameter. Next we list the relevant vertices in the non-linear 
$R_\xi$ gauge that differ from those in the linear $R_\xi$ gauge:
\begin{align}
	\begin{gathered}
		\includegraphics[scale=.2]{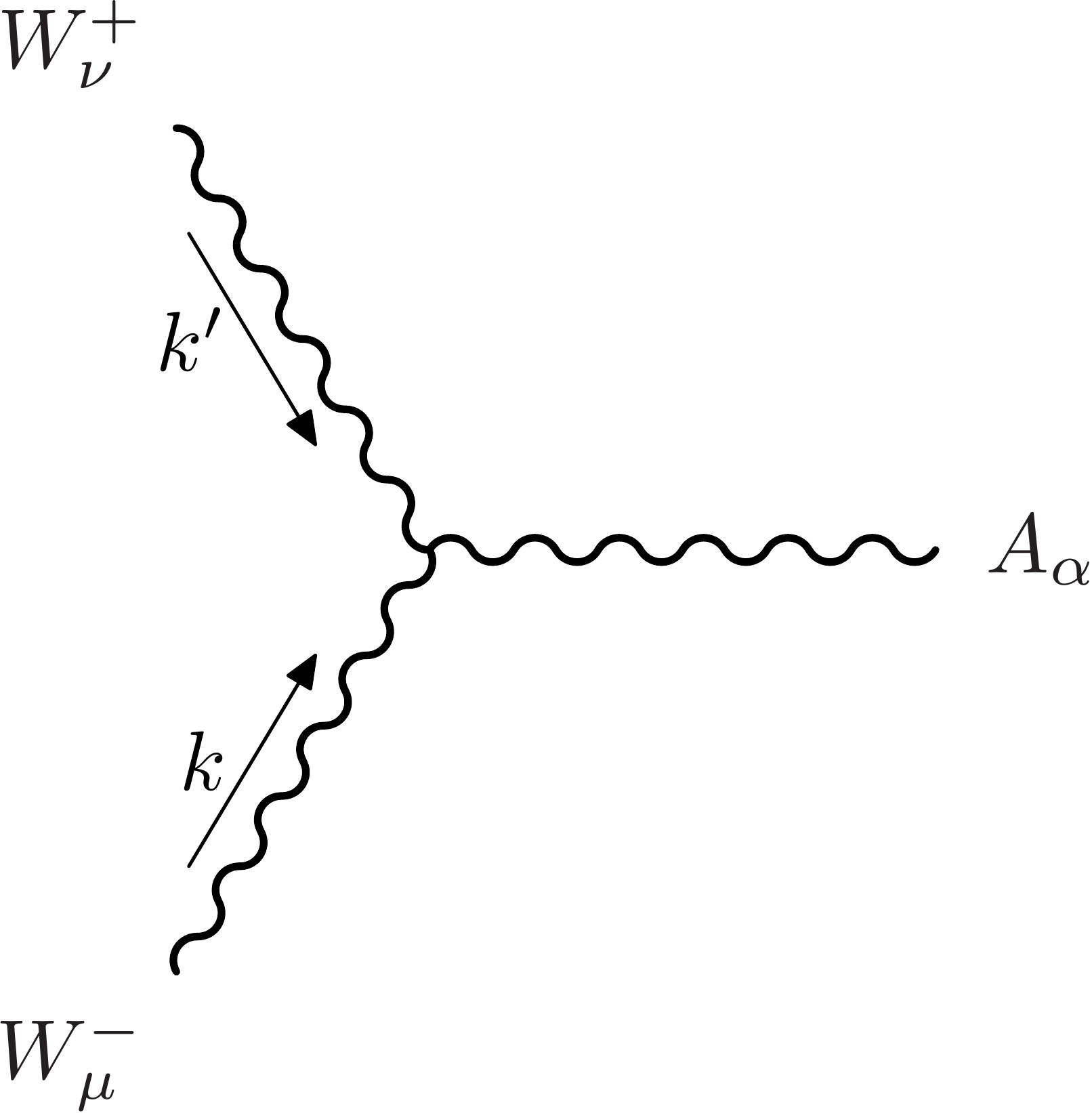}
	\end{gathered}
	\quad &=
	-ie \left[
      		g_{\mu\nu}(k-k^\prime)_\alpha 
      		- g_{\mu\alpha} 
             		 \left( 
                			2k + \left(1+ \frac{1}{\xi}\right) k^\prime 
              		\right)_\nu
      		+ g_{\nu\alpha} 
              		\left( 
                			2k^\prime + \left(1+ \frac{1}{\xi}\right) k
              		\right)_\mu
    	\right],
\end{align}
\begin{align}
	\begin{gathered}
		\includegraphics[scale=.2]{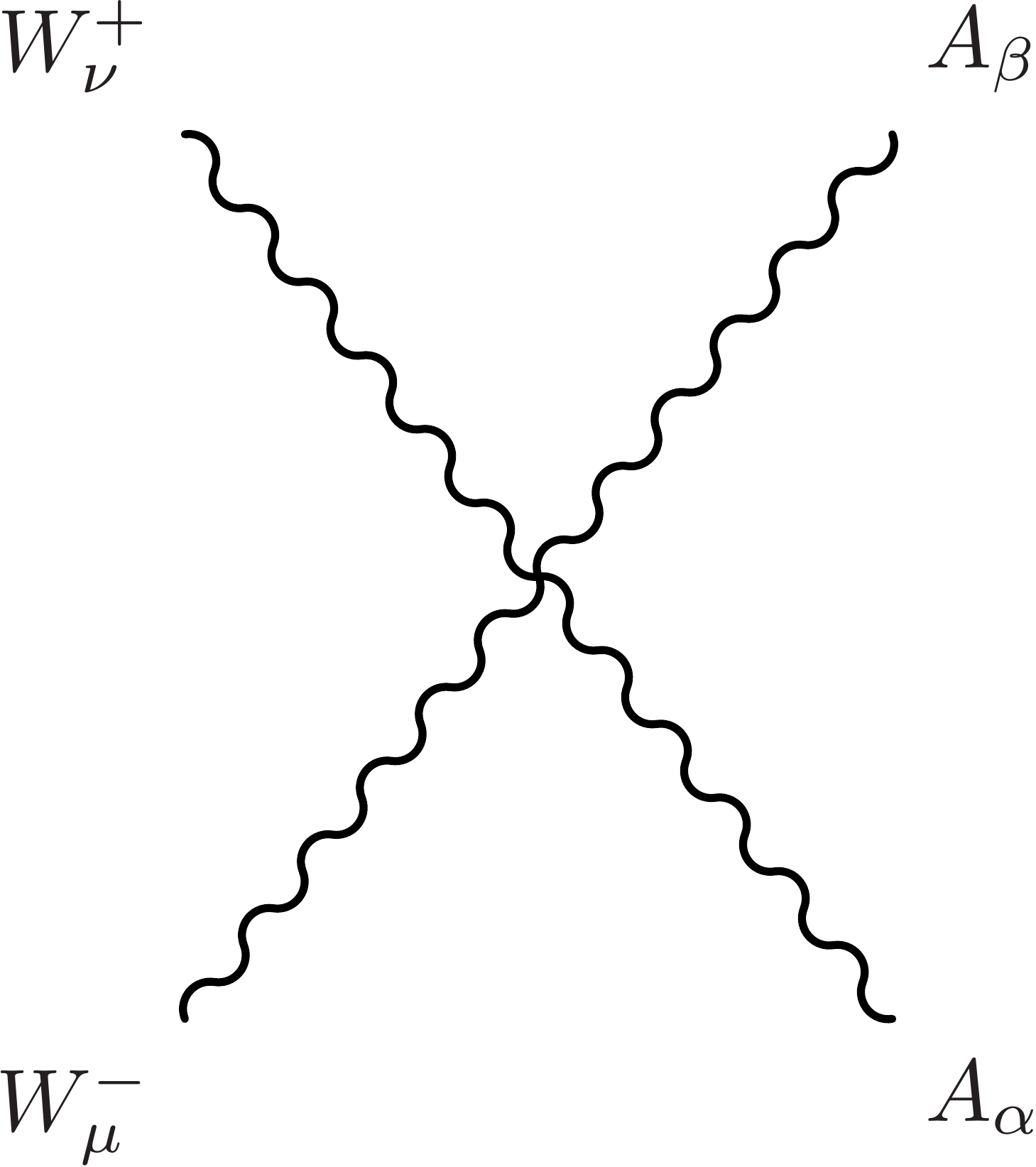}
	\end{gathered}
	\quad &=
	-ie^2 
  	\left[2g_{\mu\nu}g_{\alpha\beta}
        		-\left(1-\frac{1}{\xi}\right) 
          		\left(
            			g_{\mu\alpha}g_{\nu\beta} 
            			+ g_{\mu\beta}g_{\nu\alpha}
          		\right)
  	\right],
\end{align}
\begin{align}
	\begin{gathered}
		\includegraphics[scale=.2]{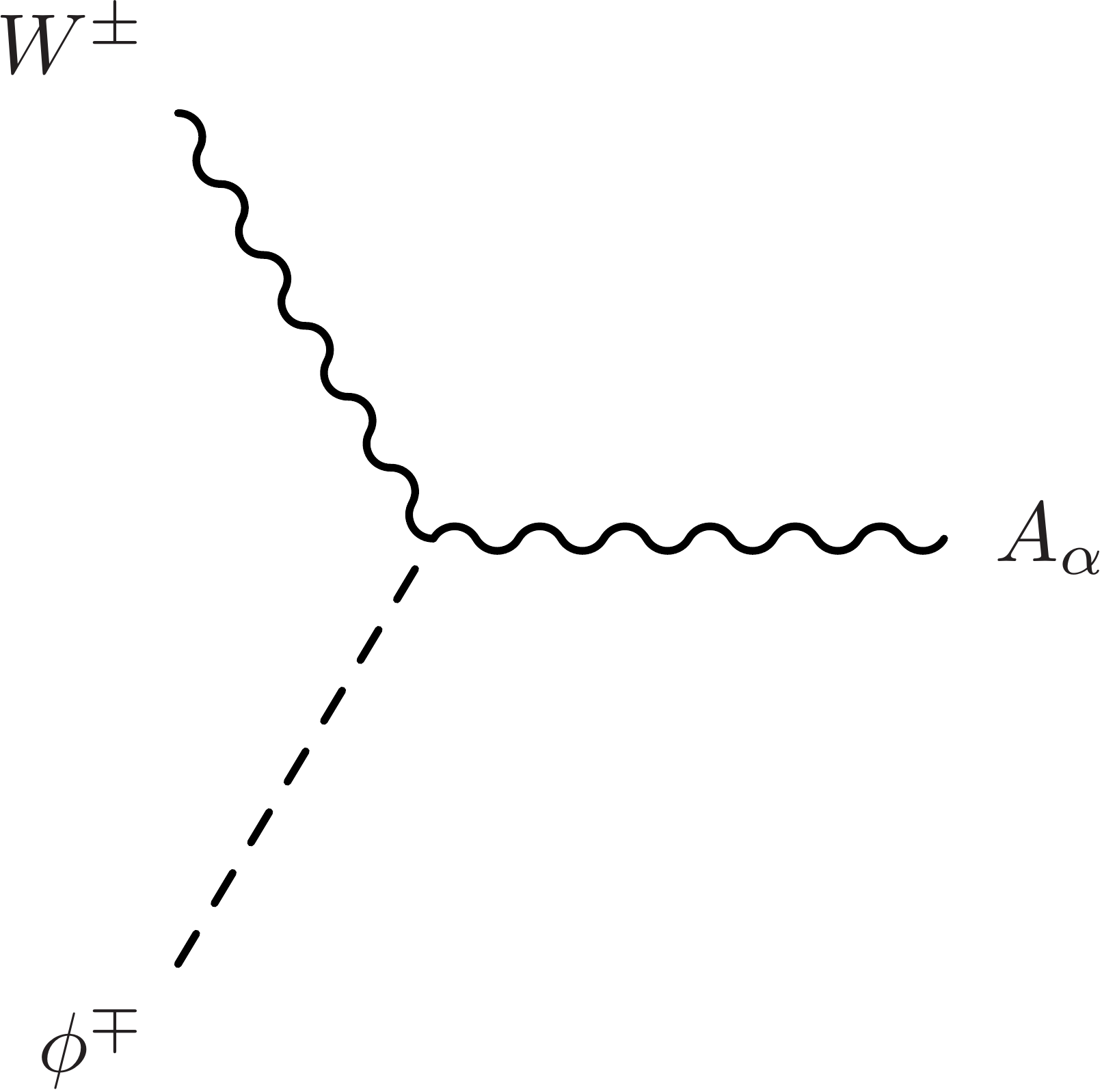}
	\end{gathered}
	\quad &= 0.
\end{align}

The remaining vertices are unchanged form those in the linear $R_\xi$ gauge 
(consult Ref.~\cite{Denner:1991kt} for a complete list of vertices in the linear $R_\xi$ gauge):
\begin{align}
	\begin{gathered}
		\includegraphics[scale=.2]{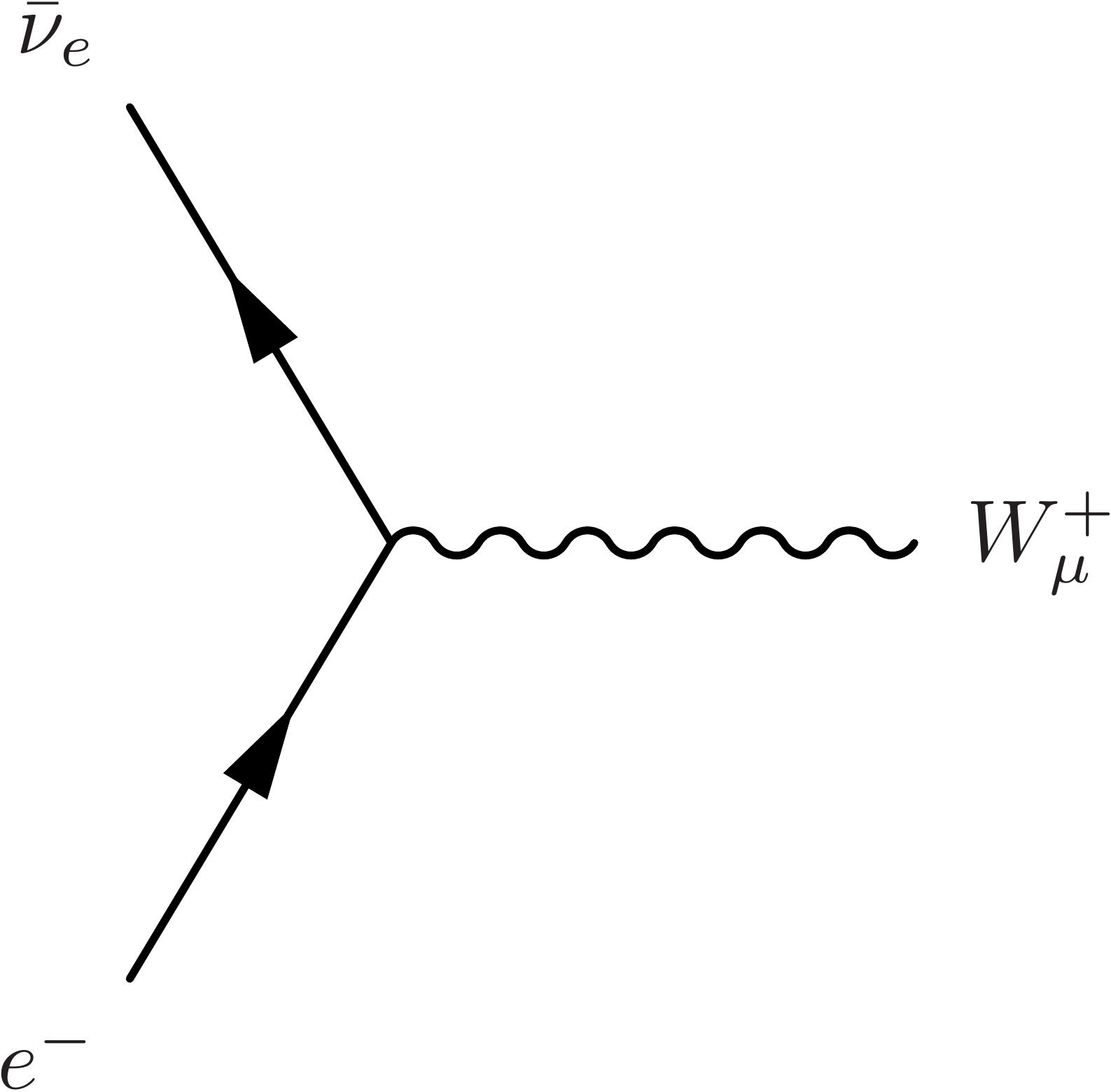}
	\end{gathered}
	\quad &= \frac{ig_W}{2\sqrt{2}} \gamma_\mu (1-\gamma_5)
	,
\end{align}
\begin{align}
	\begin{gathered}
		\includegraphics[scale=.2]{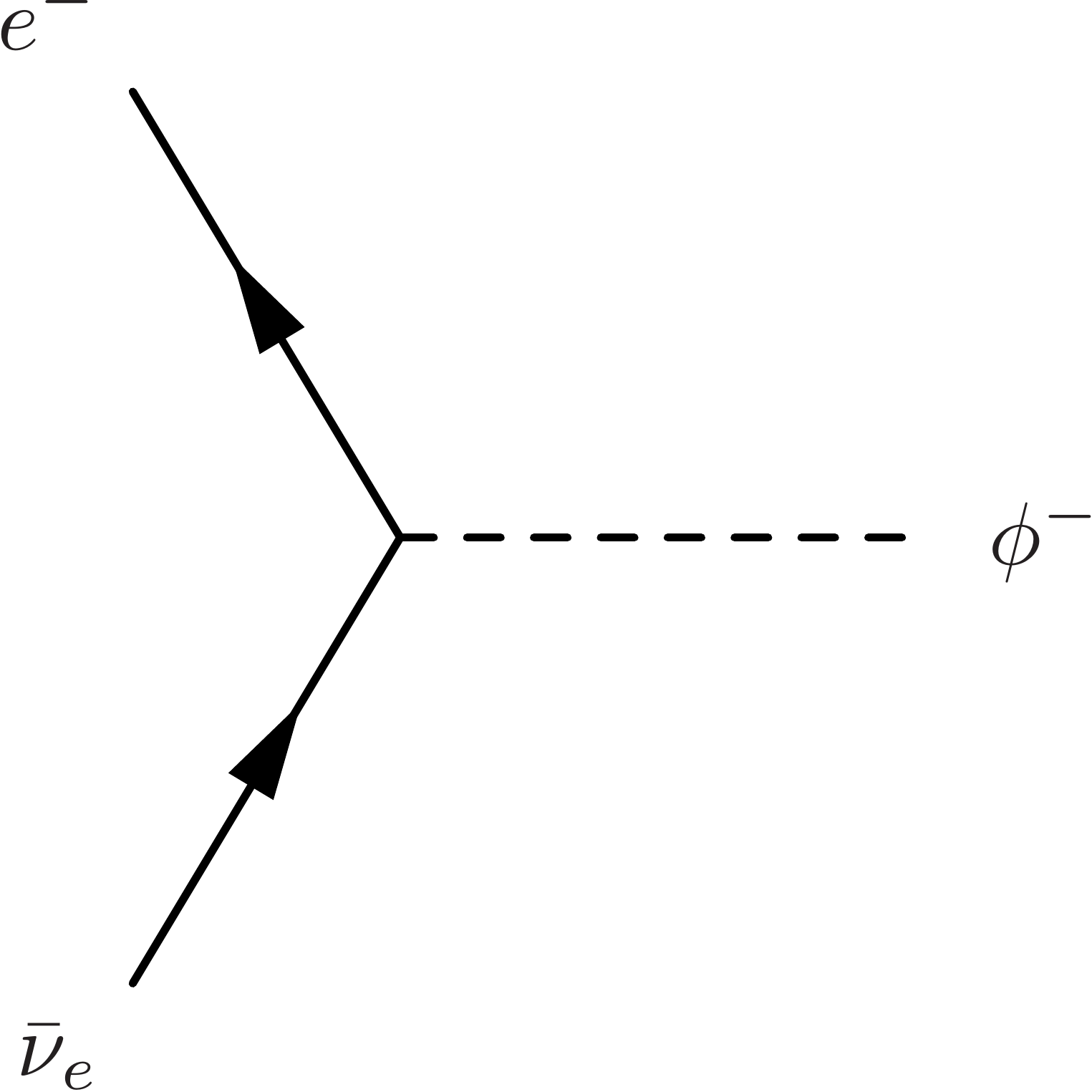}
	\end{gathered}
	\quad & = \frac{ig_W}{2\sqrt{2}} \gamma_\mu (1-\gamma_5)
	\\
	\begin{gathered}
		\includegraphics[scale=.2]{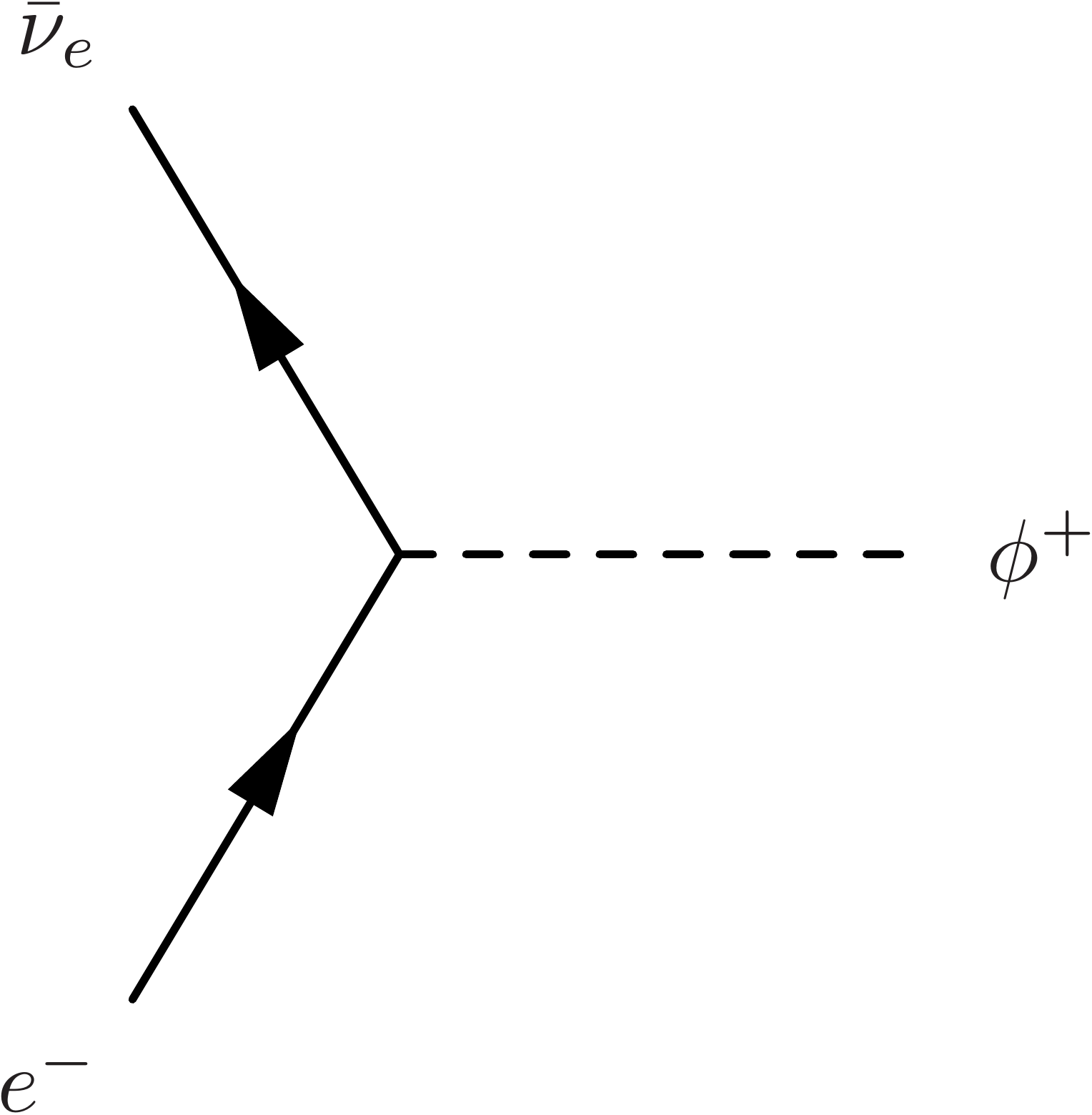}
	\end{gathered}
	\quad &= \frac{ig_W}{2\sqrt{2}} \gamma_\mu (1+\gamma_5)
	,
\end{align}
\begin{align}
	\begin{gathered}
		\includegraphics[scale=.2]{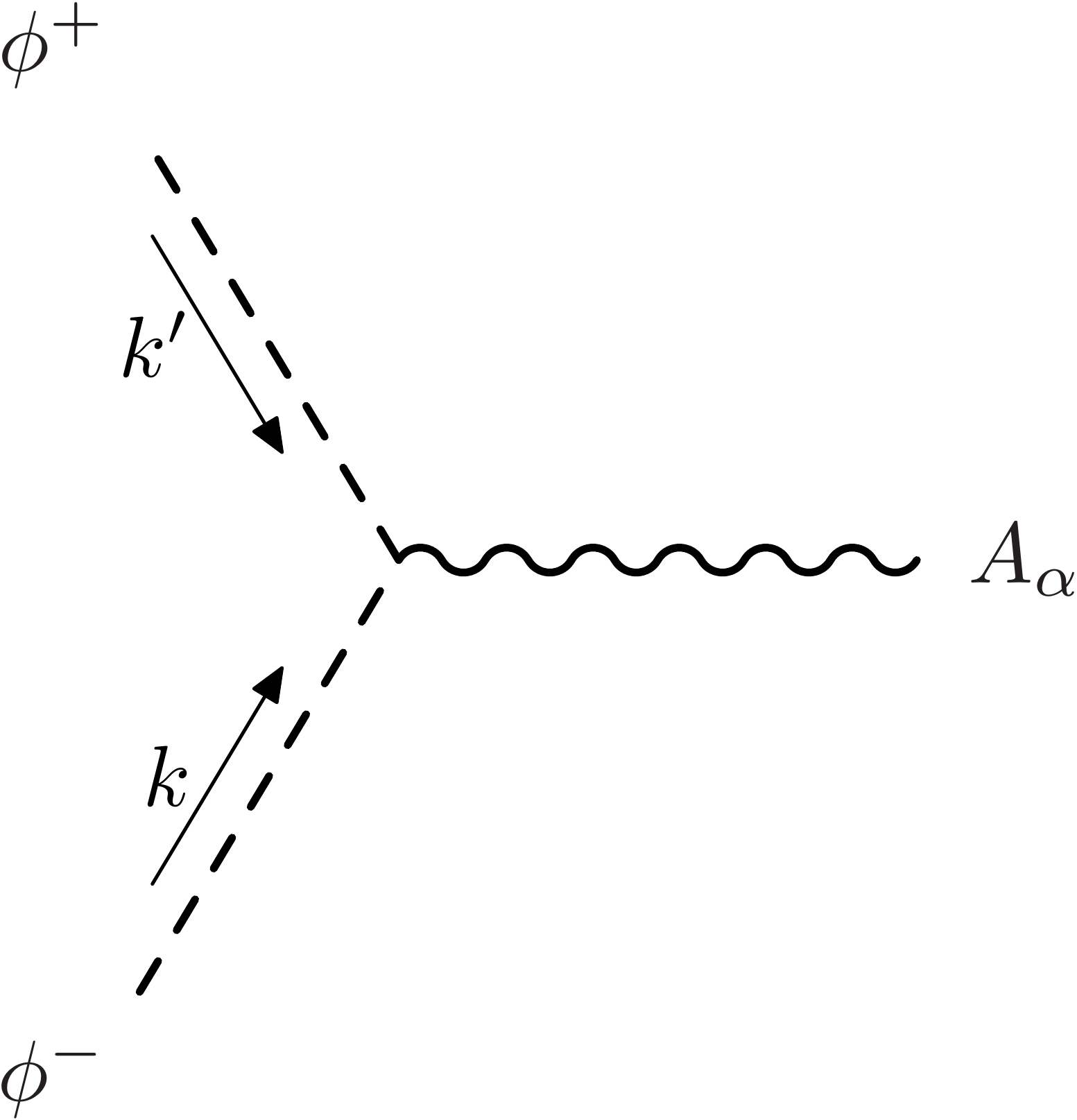}
	\end{gathered}
	\quad &= ie \left( k - k^\prime \right)_\mu
	,
\end{align}
\begin{align}
	\begin{gathered}
		\includegraphics[scale=.2]{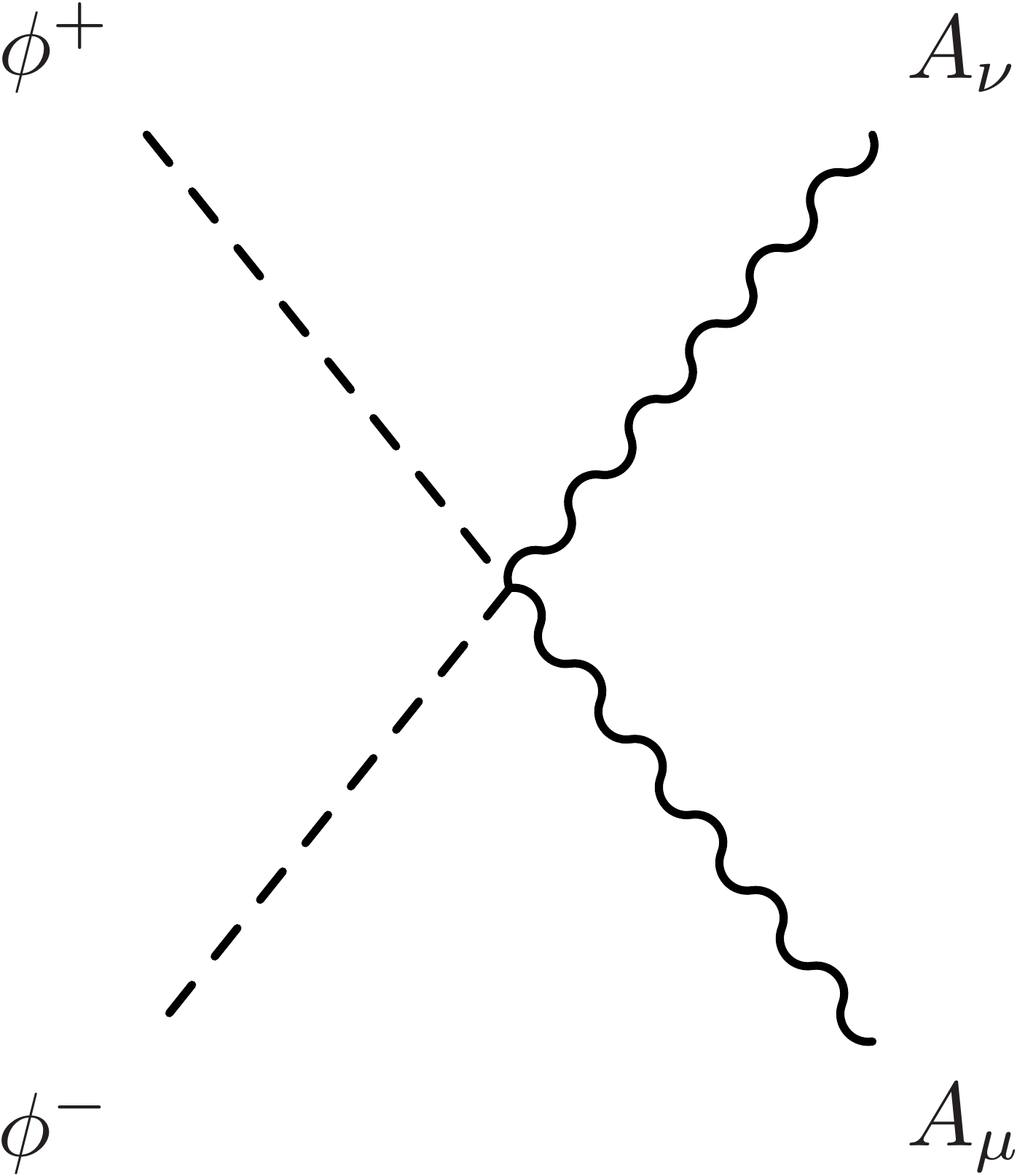}
	\end{gathered}
	\quad &= 2ie^2 g_{\mu\nu}
	.
\end{align}
Here, $g_W = e/\sin\theta_W$ is the weak coupling constant and $e>0$ is the 
electric charge of a proton.

\section{Renormalization \label{sec:renormalization}}
For our purposes, it is enough to limit our attention to axial contributions to the electron
self energy and the renormalized $e^-e^+\gamma$ vertex. The axial contributions to these 
quantities are \cite{Hollik:1993cg}
\begin{align}
	\Sigma_\text{A}(\cancel{p}) &= \cancel{p} \gamma^5 
		\left( \hat\Sigma_\text{A}(p^2) - \delta Z_\text{A} \right),
	\\
	\Gamma^\mu_\text{A} &= ie \delta Z_\text{A} \gamma^\mu \gamma^5.
\end{align} 
Above, $\hat\Sigma_\text{A}$ is the axial part of the self energy not including the counterterm. 

The counterterm contribution to the electron propagator and $e^-e^+\gamma$ vertex 
is fully determined by the axial part of the electron field strength, $\delta Z_\text{A}$.
Using on-shell renormalization conditions, the axial part of the electron field strength is
$\delta Z_\text{A} = \hat\Sigma_\text{A}(\cancel{p} = \me)$.
\end{document}